\documentstyle[12pt,fleqn]{article}
\gdef\journal#1,#2,#3,#4.{{ #1~}{\bf #2}, #3 (#4) }
\def\prd{\journal Phys. Rev. D,}
\def\prl{\journal Phys. Rev. Lett.,}

\def\npb{\journal Nucl. Phys. B,}
\def\plb{\journal Phys. Lett. B,}
\def\apj{\journal Astrophys. J.,}
\def\apjl{\journal Astrophys. J. Lett.,}
\def\MNRAS{\journal Mon. Not. R. Astron. Soc.,}
\def\be{\begin{equation}}\def\bea{\begin{eqnarray}}\def\beaa{\begin{eqnarray*}}
  \def\ee{\end{equation}}  \def\eea{\end{eqnarray}}  \def\eeaa{\end{eqnarray*}}
\def\double{\baselineskip 24pt \lineskip 10pt}

\def\la{\mathrel{\mathpalette\fun <}}

\def\fun#1#2{\lower3.6pt\vbox{\baselineskip0pt\lineskip.9pt
        \ialign{$\mathsurround=0pt#1\hfill##\hfil$\crcr#2\crcr\sim\crcr}}}
\begin{document}
\def\half{{\textstyle{ 1\over 2}}}
\def\frac#1#2{{\textstyle{#1\over #2}}}
\def\gsim{\mathrel{\raise.3ex\hbox{$>$\kern-.75em\lower1ex\hbox{$\sim$}}}}
\def\lsim{\mathrel{\raise.3ex\hbox{$<$\kern-.75em\lower1ex\hbox{$\sim$}}}}
\def\la{\bigl\langle} \def\ra{\bigr\rangle}
\def\cd{\!\cdot\!}
\def\a{\hat a}      \def\b{\hat b}      \def\c{\hat c}
\def\ab{\a\cd\b}    \def\ac{\a\cd\c}    \def\bc{\b\cd\c}
\def\cg{\cos\gamma} \def\ca{\cos\alpha} \def\cb{\cos\beta}
\def\gg{\hat\gamma}    \def\go{ \hat\gamma_1}
\def\gt{\hat\gamma_2}  \def\gth{\hat\gamma_3}
\def\got{ \hat\gamma_1\cd\hat\gamma_2} \def\ggo{ \hat\gamma\cd\hat\gamma_1}
\def\goth{\hat\gamma_1\cd\hat\gamma_3} \def\ggt{ \hat\gamma\cd\hat\gamma_2}
\def\gtth{\hat\gamma_2\cd\hat\gamma_3} \def\ggth{\hat\gamma\cd\hat\gamma_3}

\def\n{\hat n}       \def\no{\hat n_1}   \def\nt{\hat n_2}  \def\nth{\hat n_3}
\def\nont{\no\cd\nt} \def\nonth{\no\cd\nth} \def\ntnth{\nt\cd\nth}

\def\nogo{\no\cd\hat\gamma_1} \def\nogt{\no\cd\hat\gamma_2}
\def\nogth{\no\cd\hat\gamma_3}
\def\ntgo{\nt\cd\hat\gamma_1} \def\ntgt{\nt\cd\hat\gamma_2}
\def\ntgth{\nt\cd\hat\gamma_3}
\def\nthgo{\nth\cd\hat\gamma_1} \def\nthgt{\nth\cd\hat\gamma_2}
\def\nthgth{\nth\cd\hat\gamma_3}

\def\D{ {\Delta T \over T} }   \def\dO{d\Omega}

\begin{flushright}
{\footnotesize
SISSA REF. 73/94/A\\
astro-ph/9406014 }
\end{flushright}

\vspace{0.2in}

\renewcommand{\thefootnote}{\fnsymbol{footnote}}

\begin{center}
{\Large {\bf NON--GAUSSIAN EFFECTS }}

\vspace{.15in}

{\Large {\bf IN THE COSMIC MICROWAVE BACKGROUND  }}

\vspace{.15in}

{\Large {\bf FROM INFLATION}}

\vspace{.3in}

{\bf Alejandro Gangui}\footnote{Electronic mail: {\tt
gangui@tsmi19.sissa.it}}\\

\vspace{.15in}

{\em SISSA -- International School for Advanced Studies, \\
via Beirut 2 -- 4, 34013 Trieste, Italy}\\

\vspace{.37in}

To appear in {\em Physical Review D}

\end{center}

\vfill

\begin{center}
\section*{Abstract}
\end{center}

The presence of non--Gaussian features in the Cosmic Microwave Background
(CMB) radiation maps represents one of the most long--awaited clues
in the search for the actual structure of the primordial radiation,
still needing confirmation. These features could shed some light on the
non trivial task of distinguishing the real source of the primeval
perturbations leading to large scale structure. One of the simplest
non--Gaussian signals to search is the (dimensionless) skewness ${\cal S}$.
Explicit computations for ${\cal S}$ are presented in the frame of
physically motivated inflationary models (natural, intermediate and
polynomial potential inflation) in the hope of finding values in agreement
with estimated quantities from large angle scale (e.g., {\em COBE} DMR) maps.
In all the cases considered the non--Gaussian effects turn out to lie below
the level of theoretical uncertainty (cosmic variance). The possibility
of unveiling the signal for ${\cal S}$ with multiple--field models is also
discussed.

\vspace{1.4in}

PACS number(s): 98.80.Cq, 98.80.Es, 98.70.Vc, 98.80.-k

\renewcommand{\thefootnote}{\arabic{footnote}}
\addtocounter{footnote}{-1}

\newpage

\double


\vspace{18pt}
\section{Introduction}

The existence of anisotropies in the CMB radiation as recently
detected by {\em COBE} \cite{COBE92} and subsequently confirmed both
by balloon--borne scans at shorter wavelength \cite{Ganga+Co93}
and by ground--based intermediate angular scale observations
\cite{Hancock+Co94} has
triggered a large body of literature dealing with the non trivial task
of finding the correct statistics able to disentangle the relevant
information out of the primeval radiation maps.
Both small and large angle scale probes of the microwave sky have been
and are the real data our theoretical models must reproduce before we
might call them {\it viable}.
Discrimination between the two main theories for the origin of primordial
perturbations, namely, whether these are due to topological defects
\cite{DefectLect93}
produced during a GUT phase transition or to early inflationary
quantum fluctuations \cite{TurLect93}, has by now become a difficult matter.

In this paper we will
work in the frame of inflation and, in particular, we will
be mainly concerned with single--field
inflationary potentials. Some of the most popular models
are characterized by their simplicity and universality (such as quadratic and
quartic chaotic potentials), by their being exact solutions of the
equations of motion for the inflaton (like power--law  and intermediate
inflation) or by their particle physics motivation (as natural
inflation with an axion--like potential).
In contrast with these simpler models where one has just one relevant
parameter more general potentials, with more freedom, were also
considered in the literature. One example of this is the polynomial potential
\cite{Hodges+Co90} which for an adequate choice of the parameters was found
to lead to broken scale invariant spectra on a wide range of scales
with interesting consequences for large scale structure.

One should also worry about initial conditions \cite{GoldwirthPiran92}.
While for single--field models the only effect of kinetic terms
consists in slightly changing the initial value of $\phi$ (leaving
invariant the phase space of initial field values leading to sufficient
inflation), for models with more than
one scalar field initial conditions can be important,
e.g., for double inflation \cite{DoubleInflation}
(with two stages of inflation each one dominated by a different
inflaton field) leading to primordial non--Gaussian
perturbations on cosmological interesting scales.
Within the latter models, however, the question of how
probable it is that a certain initial configuration will be realized in
our neighbouring universe should be addressed.
More recently other examples of interesting multiple--field models with
broken scale invariance have also been considered
(see e.g. Ref.\cite{Salopek92}). Here all scalar fields contribute to
the energy density and
non--Gaussian features are produced when the scalar fields pass over the
interfaces of continuity of the potential.
Extension of the single--field stochastic approach developed in
Ref.\cite{3point}
for calculating the CMB angular bispectrum generated
through Sachs--Wolfe effect from primordial curvature perturbations
in the inflaton in order to include many scalar fields is therefore
needed \cite{progress}.

The plan of the paper is as follows.
In Section 2 a brief overview of some general results is given while in
Section 3 we concentrate on trying to extract numerical values for the
non--Gaussian signal (the dimensionless skewness in this case)
predicted in the frame of three different inflationary models.
Section 4 contains some general conclusions.

\section{The CMB skewness}

We will here briefly summarize the steps that lead to the calculation
of the mean two-- and three--point functions of the temperature
anisotropies and in particular to the zero--lag limit of the latter,
namely the skewness.

As usual we expand the temperature fluctuation in spherical harmonics
$\D(\vec x ;\gg) = \sum_{\ell=1}^\infty \sum_{m=-\ell}^\ell a_\ell^m (\vec x)
{\cal W}_\ell Y_\ell^m(\gg)$,
where $\vec x$ specifies the position of the observer and the unit vector
$\gg$ points in a given direction from $\vec x$.
${\cal W}_\ell$ represents the window function
of the specific experiment. Setting ${\cal W}_0 = {\cal W}_1=0$ automatically
accounts for both monopole and dipole subtraction; for $\ell \geq 2$ one can
take ${\cal W}_{\ell}\simeq\exp\left[-\frac12 \ell({\ell}+1)\sigma^2\right]$,
where $\sigma$ is the dispersion of the antenna--beam profile, which
measures the angular response of the detector (e.g. \cite{Wright+Co92}).
In some cases the quadrupole term is also subtracted from the maps
(e.g. \cite{COBE92});
in this case we also set ${\cal W}_2=0$.
The multipole coefficients $a_\ell^m$ are here considered as
zero--mean non--Gaussian random variables whose statistics
derives from that of the gravitational potential through the Sachs--Wolfe
relation $\D(\vec x ;\gg) = {1 \over 3} \Phi( \vec x + r_0 \gg)$,
where $r_0=2/H_0$ is the horizon distance and $H_0$ the Hubble constant.

In the frame of the inflationary model, the calculations reported
in Ref.\cite{3point} lead to general expressions for the
mean two-- and three--point functions of the primordial gravitational
potential, namely
$\la \Phi(r_0 \go) \Phi(r_0 \gt) \ra = (9 {\cal Q}^2 / 20\pi)
\sum_{\ell\ge 0} (2 \ell+1) P_\ell(\gg_1\cdot\gg_2) {\cal C}_\ell$
and
\bea
\label{eq717}
\lefteqn{
\la \Phi(r_0 \go) \Phi(r_0 \gt) \Phi(r_0 \gth) \ra
= {81\pi^2 {\cal Q}^4 \over 25 (2\pi)^4} \Phi_3
\sum_{j, \ell \ge 0} (2j+1)(2\ell+1) {\cal C}_j {\cal C}_\ell
	}
\nonumber \\
&  &
\times
\bigl[ P_j(\go \cdot \gth) P_\ell(\go \cdot \gt) +
       P_j(\gt \cdot \go) P_\ell(\gt \cdot \gth) +
       P_j(\gth \cdot \go) P_\ell(\gth \cdot \gt) \bigr] \ ,
\eea
with $P_\ell$ a Legendre polynomial and
where $\Phi_3$ is a model--dependent coefficient.
These expectation values are a statistical average over the ensemble of
possible observers and can only depend upon the needed number of
angular separations.\\
The $\ell$--dependent coefficients ${\cal C}_\ell$ are
defined by $\la Q_\ell^2 \ra \equiv {(2 \ell + 1)  \over 5} {\cal Q}^2
{\cal C}_\ell$, with ${\cal Q} = \la Q_2^2 \ra^{1/2}$ the {\em rms}
quadrupole, and are
related to the gravitational potential power--spectrum $P_\Phi(k)$ through
${\cal C}_\ell = \int_0^\infty dk k^2 P_\Phi(k) j_\ell^2(k r_0) /
\int_0^\infty dk k^2 P_\Phi(k) j_2^2(k r_0)$,
where $j_\ell$ is the $\ell$--th order spherical Bessel
function.
The {\em rms} quadrupole is simply related to the quantity
$Q_{rms-PS}$ defined in Ref.\cite{COBE92,Bennett+Co94}: ${\cal Q} =\sqrt{4\pi}
Q_{rms-PS}/T_0$, with mean temperature $T_0 = 2.726\pm 0.01 K$
\cite{Mather+Co93}. For the scales of interest we can make the
approximation $P_\Phi(k) \propto k^{n-4}$, where $n$ corresponds to the
primordial index of density fluctuations (e.g. $n=1$ is the Zel'dovich,
scale--invariant case), in which case \cite{BE87,FLM87} we have
${\cal C}_\ell =
 \Gamma\left(\ell+\frac{n}{2}-\frac{1}{2}\right)
 \Gamma\left(\frac{9}{2}-\frac{n}{2}\right)
 /
\left(
 \Gamma\left(\ell+\frac{5}{2}-\frac{n}{2}\right)
 \Gamma\left(\frac{3}{2}+\frac{n}{2}\right)
\right)$.
The equations above allow us to compute the angular spectrum,
$\la a_{\ell_1}^{m_1} {a_{\ell_2}^{m_2}}^* \ra =
\delta_{\ell_1\ell_2} \delta_{m_1 m_2}{\cal Q}^2{\cal C}_{\ell_1} / 5$,
and the angular bispectrum,
\be
\la a_{\ell_1}^{m_1} a_{\ell_2}^{m_2} {a_{\ell_3}^{m_3}}^* \ra
= {3 {\cal Q}^4 \over 25} \Phi_3 \bigl[ {\cal C}_{\ell_1} {\cal C}_{\ell_2} +
{\cal C}_{\ell_2} {\cal C}_{\ell_3} +
{\cal C}_{\ell_3} {\cal C}_{\ell_1}
\bigr] {\cal H}_{\ell_3 \ell_1 \ell_2}^{m_3 m_1 m_2} \ ,
\ee
where the coefficients ${\cal H}_{\ell_1\ell_2\ell_3}^{m_1 m_2 m_3}
\equiv \int \dO_{\gg} {Y_{\ell_1}^{m_1}}^*(\gg) Y_{\ell_2}^{m_2}(\gg)
Y_{\ell_3}^{m_3}(\gg)$
are only non--zero if
the indices $\ell_i$, $m_i$ ($i=1,2,3$) fulfill the relations:
$\vert \ell_j - \ell_k \vert \leq  \ell_i \leq \vert \ell_j + \ell_k \vert$,
$\ell_1 + \ell_2 + \ell_3 = even$ and $m_1 = m_2 + m_3$.
{}From the last equation we obtain the general
form of the mean three--point correlation function for the temperature
perturbations whose explicit form we
will not need in the present work. Some simplifications occur for the
CMB mean skewness $\la C_3(0) \ra$ for which we obtain
\bea
\label{eq57}
\lefteqn{
\la C_3 (0) \ra = {3 {\cal Q}^4 \over 25 (4\pi)^2} \Phi_3
\sum_{\ell_1,\ell_2,\ell_3}\!\! (2 \ell_1 +1) (2 \ell_2 +1) (2 \ell_3 +1)
\bigl[ {\cal C}_{\ell_1} {\cal C}_{\ell_2} +
{\cal C}_{\ell_2} {\cal C}_{\ell_3} +
{\cal C}_{\ell_3} {\cal C}_{\ell_1} \bigr]
	}
\nonumber \\
&  &
\times
{\cal W}_{\ell_1} {\cal W}_{\ell_2} {\cal W}_{\ell_3}
{\cal F}_{\ell_1 \ell_2 \ell_3}
\eea
where the coefficients ${\cal F}_{\ell_1 \ell_2 \ell_3} \equiv
(4\pi)^{-2}\int\dO_{\gg} \int\dO_{\gg'} P_{\ell_1}(\gg\cdot\gg') P_{\ell_2}
(\gg\cdot\gg') P_{\ell_3}(\gg\cdot\gg')$ may be suitably expressed in terms of
products of factorials of $\ell_1$, $\ell_2$ and $\ell_3$, using standard
relations for Clebsch--Gordan coefficients.

By working in the frame of the stochastic approach to inflation
\cite{Starobinskii86,Goncharov+Co87} we are able to compute the
three--point function for the inflaton field perturbation $\delta \phi$
\cite{3point}.
Among the primordial perturbation scales we find those that
stretched to super--horizon sizes approximately 60 e--foldings before the
end of inflationary era to re--enter the Hubble radius next,
during matter domination, as energy--density perturbations (and thus
affecting also the gravitational potential $\Phi$).
During decoupling
these scales are still greater than the horizon and
therefore no microphysics alters the primevally imprinted signal they
carry.
The absence of non--linear evolution for these large angle scales makes
physics simple and therefore highly predictive.

The stochastic analysis naturally takes into account all the multiplicative
effects in the
inflaton dynamics that are responsible for the non--Gaussian features.
Extra--contributions to the three--point
function of the gravitational potential also arise as a consequence of
the non--linear
relation between $\Phi$ and $\delta \phi$ \cite{BarrowColes90}.
When all primordial second--order effects are taken into account
we get
\be
\label{quad}
{\cal Q}^2 = {8 \pi^2 H_{60}^2 \over 5 m_P^2 X_{60}^2}
{\Gamma (3-n) \Gamma\left(\frac{3}{2}+\frac{n}{2}\right)
\over \left[\Gamma\left(2-\frac{n}{2}\right)\right]^2
\Gamma (\frac{9}{2}-\frac{n}{2})}
\ee
for the {\em rms} quadrupole,
while for the ``dimensionless" skewness
${\cal S} \equiv \la C_3(0) \ra / \la C_2 (0)\ra ^{3/2}$
we find \cite{f1}
\be
\label{skew}
{\cal S} =
{\sqrt{45\pi}\over 32 \pi^2} {\cal Q}
\left[X_{60}^2 - 4 m_P X'_{60} \right] {\cal I}(n)
\nonumber
\ee
where we denoted with $X_{60}$ the value of the steepness of the
potential $X(\phi) = m_P V'(\phi) / V(\phi)$ evaluated at $\phi_{60}$
(the value of the inflaton 60 e--foldings before the end of inflation)
and
where ${\cal I}(n)$ is a spectral index--dependent geometrical factor
of order unity for interesting values of $n$. \cite{3point}

It is also of interest here to calculate the accurate form of the
spectrum of primordial perturbations (e.g. by finding the value of the
spectral index) at the moment the scales relevant for our study left
the Hubble radius.
Our expression ${\cal S}$ for the dimensionless skewness is accurate
up to second order in perturbation theory (the first non vanishing
order) and therefore we
should calculate $n$ at least to the same order.
We will borrow the notation for
the slow--roll expansion parameters from
Ref.\cite{KolbVadas94}.
These are defined as
$\epsilon(\phi) = m_P^2 \left( H'(\phi)/ H(\phi) \right)^2 / 4 \pi $,
$\eta(\phi)  =  m_P^2  H''(\phi) / (4\pi H(\phi) ) $ and
$\xi(\phi)  = m_P^2 H'''(\phi) / ( 4\pi H'(\phi))$ \cite{f2}.
In the slow-roll approximation $\epsilon$ and $\eta$ are less than one.
The same is not true in general for $\xi$ and this may cause consistency
problems when this term is incorrectly neglected
(see the discussion in \cite{KolbVadas94}).
We will see below examples of this.

Let ${\cal Q}_S^2$ (${\cal Q}_T^2$) be the contribution of the scalar
(tensor) perturbation to the variance of the quadrupole temperature
anisotropy.  The complete second--order expressions for the tensor
to scalar ratio and for the spectral index are given respectively by
$R \equiv {\cal Q}_T^2 / {\cal Q}_S^2
\simeq 14\epsilon\left[1-2C(\eta-\epsilon )\right]$
and
\be
\label{12}
n  = 1 -  2\epsilon\left[2-\frac{\eta}{\epsilon}+4(C+1)\epsilon -
(5C+3)\eta + C\xi (1+2(C+1)\epsilon - C\eta) \right],
\ee
evaluated at $\phi\simeq\phi_{60}$.
In this equation
$C\equiv -2+\ln 2 +\gamma\simeq -0.7296$ and $\gamma=0.577$ is
the Euler--Mascheroni constant \cite{StewartLyth93}.

It was realized \cite{SV91,AbbWise84,Srednicki93}
that a positive detection of a non--zero
three--point function (an in particular a non--zero skewness)
in the temperature fluctuations on the microwave sky
{\em does not} imply an intrinsically non--Gaussian underlying field.
This problem is related to the so--called ``cosmic variance".
Relevant for large angular scale fluctuations, this variance is in fact
the strongest source of (theoretical) noise we have to deal with and
essentially reflects the impossibility of making observations in more
than one universe.
One way to quantify this effect is through the {\em rms} skewness
of a Gaussian field  $\la C_3^2(0) \ra _{Gauss}^{1/2}$, which we may
express as
$\la C_3^2(0) \ra _{Gauss} = 3\int^1_{-1}d\cos\alpha\la C_2(\alpha)\ra^3$.
A convenient quantity to compare with ${\cal S}$ is
the normalized {\it rms} skewness
$\la C_3^2(0) \ra_{Gauss}^{1/2} / \la C_2(0)\ra ^{3/2}$.
For interesting values of $n$ this ratio is of the order $\sim 0.1$.
A rough criterion for the feasibility of detecting
primordial non--Gaussian signatures could be expressed as
$\la C_3(0) \ra \gsim  \la C_3^2(0) \ra_{Gauss}^{1/2}$.
Unfortunately as we will see
for the models we work with here this is far from being the
case; as a result primordial features cannot emerge.
It is worth mentioning that recent analyses of the
three--point function and the skewness from {\em COBE} data
\cite{Hinshaw+Co93,Smoot+Co93} are also consistent with
quasi--Gaussian fluctuations.

Now let us turn to the examples.

\section{Worked Examples}

\vspace{18pt}

\subsection{Natural Inflation}

To begin with let us consider the Natural inflationary scenario. First
introduced in \cite{Natural} this model borrows speculative ideas from
axion particle physics \cite{KT90}. Here the existence of disparate mass
scales leads to the explanation of why it is physically attainable to
have potentials with a height many orders of magnitude below its width
\cite{Adams+Co91},
as required for successful inflation where usually self coupling
constants are fine--tuned to very small values.

This model considers a Nambu-Goldstone (N-G)
boson, as arising from a spontaneous symmetry breakdown of a global
symmetry at energy scale $f\sim m_P$, playing the role of the inflaton.
Assuming there is an additional explicit symmetry breaking phase at mass scale
$\Lambda\sim m_{GUT}$ these particles become pseudo N-B bosons and a periodic
potential due to instanton effects arises.
The simple potential (for temperatures $T\leq\Lambda$) is of the form
$V(\phi) = \Lambda^4 [1+\cos(\phi / f)]$.

For us this $V(\phi)$ constitutes an axion--like
model with the scales $f$ and $\Lambda$ as free parameters.
It is convenient to split the parameter space into two regions.
In the $f >> m_P$ zone the whole inflationary period happens in the
neighbourhood of the minimum of the potential, as may be clearly seen
from the slow--rolling equation \cite{ST84}
$ \vert\cos (\phi / f)\vert \leq 24 \pi / (24 \pi + (m_P / f)^2 ) $
which is only violated near $\phi_{end} \simeq \pi f$, and from the small
value of the steepness
$\vert X \vert = {m_P \over f } \tan ({\phi\over 2 f})$
for $\phi$ smaller than $\phi_{end}$ \cite{Turner93}.
Thus by expanding $V(\phi)$ around the minimum it is easy to see the
equivalence between this potential and the quadratic one
$V \sim m^2 (\phi - \pi f)^2 / 2$ with $m^2 = \Lambda^4 / f^2$.
We have already studied the latter in Ref.\cite{3point} and we will add
nothing else here.
On the other hand, let us consider the other regime,
where $f \lsim m_P$. Reheating temperature considerations place a lower
limit on the width $f$ of the potential. For typical values of the model
parameters involved, a temperature $T_{RH} \lsim 10^8$GeV is attained.
GUT baryogenesis via the usual out--of--equilibrium decay of X--bosons
necessitates instead roughly $T_{RH} \sim 10^{14}$GeV
(the mass of the gauge bosons) for successful reheating \cite{ST84}.
Thus the final temperature is not high enough to create them from the
thermal bath.
Baryon-violating decays of the field and its products
could be an alternative to generate the observed asymmetry if taking
place at $T_{RH} > 100$GeV, the electroweak scale.
This yields the constraint $f \gsim 0.3 m_P$ \cite{FreeseConf93}, implying
$n\gsim 0.6$ (see below).
Attractive features of this model include the possibility of having a
density fluctuation spectrum with extra power on large scales. Actually
for $f\lsim 0.75~m_P$ the spectral index may be accurately expressed as
$n\simeq 1 - m_P^2 / 8\pi f^2$ \cite{f3}.
This tilt in the spectrum as well as the negligible
gravitational wave mode contribution to the CMB anisotropy might lead to
important implications for large scale structure.\cite{Adams+Co93}

Let us take now $f\simeq 0.446~m_P$ corresponding to $n\simeq 0.8$.
Slow--rolling requirements are satisfied provided the accelerated
expansion ends by $\phi_{end}\simeq 2.78 f$, very near the minimum of the
potential. Furthermore, the slow--rolling solution of the field equations
yields
the value of the scalar field 60 e-foldings before the end
of inflation
$\sin(\phi_{60}/2 f) \simeq \exp(- 15 m_P^2 / 4\pi f^2 )$
where we approximated $\phi_{end} \simeq \pi f$.\\
We find
$ X_{60}^2 - 4 m_P X_{60}' = (m_P / f)^2 [2+\sin^2(\phi_{60} / 2 f)]
                            (1-\sin^2(\phi_{60} / 2 f) )^{-1}
                            \simeq 2 (m_P / f)^2 $.

Two years of data by {\em COBE}
\cite{Bennett+Co94} are not yet enough to separately
constrain the amplitude of the quadrupole and the spectral index. A
maximum likelihood analysis yields
$Q_{rms-PS} = 17.6 \exp[0.58 (1-n)]\mu K$.
By making use of Eq.(\ref{quad}) for the {\it rms} quadrupole, {\em COBE}
results constrain the value of the free parameter
$\Lambda \simeq 1.41\times 10^{-4} m_P$.
We find ${\cal S}\simeq 3.9\times 10^{-5}$,
a rather small signal for the non--Gaussian amplitude of the fluctuations.

\subsection{Intermediate Inflation}

We will now study a class of universe models where the scale factor
increases at a rate intermediate between power--law inflation --as
produced by a scalar field with exponential potential
\cite{PLI85} -- and the standard de Sitter inflation.
In Ref.\cite{Barrow90} Barrow shows that it is possible to parameterize these
solutions by an equation of state with pressure $p$ and energy density
$\rho$ related
by $\rho + p = \gamma \rho^{\lambda}$, with $\gamma$ and $\lambda$
constants.
The standard perfect fluid relation is recovered for $\lambda = 1$
leading to the $a(t)\sim e^{H_{inf} t}$ ($H_{inf}$ constant during
inflation) solution of the
dynamical equations when the spatial curvature $k = 0$ and $\gamma = 0$,
while  $a(t)\sim t^{2/3\gamma}$ for $0<\gamma< 2/3$.
This non--linear equation of state (and consequently the two limiting
accelerated expansion behaviours) can be derived from a scalar field with
potential $V = V_0 \exp (- \sqrt{3\gamma}\phi)$.
On the other hand for $\lambda > 1$ we have
$a(t)\propto \exp (A t^f)$ (intermediate inflation) with $A > 0$ and
$0 < f \equiv 2 (1 - \lambda) / (1 - 2\lambda) < 1$ and again in this
last case it is possible to mimic the matter source with that produced
by a scalar field $\phi$, this time with potential
\be
\label{IntPot}
V(\phi) =
{8 A^2\over (\beta + 4)^2} \left({m_P\over \sqrt{8\pi}}\right)^{2+\beta}
\left({\phi\over \sqrt{2 A \beta}}\right)^{-\beta}
\left(
6 - {\beta^2 m_P^2 \over 8\pi}\phi^{-2}
\right)
\ee
with $\beta = 4 (f^{-1} - 1)$.

The equations of motion for the field in a $k = 0$ Friedmann universe
may be expressed by
$3 H^2 = 8\pi ( V(\phi) + \dot\phi^2 /2 ) / m_P^2 $
and
$\ddot\phi + 3 H \dot\phi = - V'$.
Exact solutions for these equations with the potential of Eq.(\ref{IntPot})
are of the form \cite{Barrow90,BarrowLiddle93}:
$H(\phi) = A f (A \beta / 4\pi )^{\beta /4} (\phi /m_P)^{-\beta /2}$
and
$\phi(t) = (A \beta / 4\pi )^{1/2} t^{f/2} m_P$.
Solutions are found for all $\phi > 0$ but only for
$\phi^2 > (\beta^2 / 16 \pi) m_P^2$
we get $\ddot a > 0$ (i.e. inflation).
In addition $\beta > 1$ is required to ensure that the accelerated
expansion occurs while the scalar field rolls (not necessarily
slowly) down the potential, in the region to the right of the maximum
(as it is generally the case).

{}From the full potential (\ref{IntPot}) we may compute the value of
the dimensionless skewness ${\cal S}$ (for convenience we will be taking
the field $\phi$ normalized in Planck mass units from here on)
\be
\label{11}
{\cal S} =
0.17 {\cal Q}   \left[
{\beta (\beta - 4)\over \phi^2} + {4\beta^4 (\beta - 1) -
192\pi \beta^2 (\beta - 6)\phi^2 \over \phi^2 (48\pi \phi^2 - \beta^2)^2 }
\right] ,
\ee
evaluated at $\phi\simeq\phi_{60}$.
In Eq.(\ref{11}) we took ${\cal I}(n)\simeq 4.5$ (from
Eq.(\ref{skew})), that is the case for the specific examples we discuss
below. A plot of this quantity as a function of
$d\equiv \phi^2 - \beta^2 / 16\pi > 0$
(the value of the squared of the field beyond the minimum allowed)
for different values of $\beta$ is
given in Fig. 1. Note that both positive and negative values of
${\cal S}$ are therefore allowed just by modifying the choice of
$\beta$. Another generic feature is the rapid decrease of the
non--Gaussian amplitude for increasing values of the field beyond
$\beta / \sqrt{16\pi}$ (i.e. $d>0$). Clearly this is because
for large $\phi$ we approach the slow--roll region where the steepness
becomes increasingly small.

Similar calculations may be done for the spectral index.
The explicit expression of $n$ calculated from our potential (\ref{IntPot})
is complicated and not very illuminating.
Fig. 2   illustrates the variation of $n$ as a function of $d$
(i.e. the scale dependence of the spectral index)
for different values of the parameter $\beta$.

These two figures show that for acceptable values of the spectral index
very small amplitudes for ${\cal S}$ are generally predicted.
As an example we consider $\beta = 1.2$. This ansatz yields a negative
${\cal S}$ being $\phi_{60} \simeq 0.37$ the value of the field that
maximizes $\vert {\cal S}\vert \simeq 2.25 {\cal Q}$. We show in Fig. 3
the form of the inflaton potential (\ref{IntPot}) for this particular
$\beta$. We see that for the scales that exit the Hubble radius 60
e--foldings before the end of inflation \cite{f4}
the value of the inflaton, $\phi_{60}$, is located in the steep region
beyond the maximum of the potential.
The value of the spectral index associated with this choice of $\beta$ is
$n\simeq 1.29$ (a 3\% below the first--order result $n\simeq 1.34$).
This value in excess of unity for the scales under
consideration yields a spectrum with less power on large scales
(compared with a Harrison--Zel'dovich one) making the long wave length
gravitational wave contribution to the estimated quadrupole subdominant.
The slow-roll parameters
for this scale are $\epsilon = 0.16$, $\eta = 0.37$ and $\xi = 1.01$
\cite{f5}.
While the first two are smaller than one, $\xi$ is not and so cannot be
considered an expansion variable on the same footing as $\epsilon$ and
$\eta$. Terms proportional to $\xi$ (and therefore non negligible)
in Eq.(\ref{12}) are those not included in the second--order analysis done for
the first time in Ref.\cite{StewartLyth93}.
Taking the {\em COBE} normalization we finally get
${\cal S} \simeq - 4.4\times 10^{-5}$.

Let us now consider $\beta = 7$. Now the potential falls to zero as a
power--law much more rapidly than in the previous case.
If we take $n\simeq 0.8$ we see from Fig. 2 that $d\simeq 6.05$
($\phi_{60}\simeq 2.65$) corresponding to the
slow--rolling region of the potential \cite{f6}.
For the parameters we find the following  values: $\epsilon = 0.13$,
$\eta = 0.17$ and $\xi = 0.27$. From these we see that the first--order
result ($n\simeq 0.86$)  is 7 \% above the full second--order one.
In this case we get
${\cal S} \simeq 0.50 {\cal Q}$.
Now gravitational waves contribute substantially to the detected
quadrupole. Actually we have
${\cal Q}_T^2 / {\cal Q}_S^2 \simeq 1.99$
\cite{LiddleLyth93,BarrowLiddle93,KolbVadas94}
(while we would have had $\sim 1.89$ up to first order).
Thus the estimated quadrupole should be multiplied by a factor
$(1 + 1.99)^{-1/2}$ to correctly account for the tensor mode
contribution. Finally we get
${\cal S} \simeq  7.5\times 10^{-6}$.

\subsection{Polynomial Potential}

We are interested in considering a potential of the form
\be
\label{PolyPot}
V(\phi) = A ({1\over 4}\phi^4 + {\alpha\over 3}\phi^3 + {\beta\over
8}\phi^2 ) + V_0
\ee
where for convenience $\phi$ is written in Plack mass units and $A$,
$\alpha$ and $\beta$ are dimensionless parameters. Translation
invariance allows us to omit the linear $\phi$ contribution to $V$.
A detailed analysis of a potential of the form (\ref{PolyPot}) was done
by Hodges et al. \cite{Hodges+Co90}.
Parameter space diagrams were constructed and
regions where non--scale invariance was expected were isolated. Here we
will just summarize what is necessary for our study.

Taking $\beta > 8\alpha^2 / 9$ ensures that $\phi = 0$ is the global
minimum and therefore $V_0 = 0$.
If we further require $\beta > \alpha^2 $ then no false vacua are
present.
Scalar curvature perturbations are conveniently expressed in terms of the
gauge--invariant variable $\zeta$ \cite{BST83,LiddleLyth93}.
The power spectrum associated with it, assuming slow--roll evolution of
the scalar field in the relevant region of the potential, is given by
$P_{\zeta}^{1/2}(k)\propto H^2 / \dot\phi $ evaluated at horizon crossing
time.
Equivalently $P_{\zeta}^{1/2}(k)\propto V^{3/2} / (m_P^3 V') $ which
suggests that regions of the parameter space  where the slope of the
potential goes through  a minimum or a maximum will be of interest as far
as broken scale invariance is concerned.
The presence of these extrema in $V'$ is guaranteed by taking
$\alpha^2 < \beta < 4\alpha^2 / 3$, with $\alpha < 0$ in order for the
scalar field to roll down the potential from the right.
We will thus concentrate our study in the vicinity of  an inflection
point $\phi_f$, i.e. near the curve $\beta = \alpha^2$ (but still
$\beta > \alpha^2$). In this limit we have $\phi_f \simeq -{\alpha\over 2}$
and the number of e-foldings taking place in the region of approximately
constant slope about $\phi_f$ is given by
$N = -\pi\alpha^3 / \sqrt{3(\beta - \alpha^2)}$ \cite{Hodges+Co90}.
Fixing $N=60$ as the number required for achieving sufficient inflation,
the interesting parameters ought to lie close to
the curve
$\beta \simeq \alpha^2 + \pi^2 \alpha^6 / (3 N^2)
       \simeq \alpha^2 + y \alpha^6 $
with $y \simeq 9.1\times 10^{-4}$.

We find
$ X_{60}^2 - 4 X_{60}' = 192~ y ~(9 y - \alpha^{-4}) /
[\alpha^2(6 y + \alpha^{-4})^2]$
where, again, we are normalizing the field in Planck mass
units, $y \simeq 9.1\times 10^{-4}$ and we have
evaluated the field for $\phi_{60} = \phi_f\simeq -{\alpha\over 2}$.

Variation of $\alpha$ in the allowed range results in both
positive and negative values for ${\cal S}$. Clearly
$\beta > 10\alpha^2 / 9$ yields ${\cal S} > 0$.
Although in this region of the parameter space the value of $\alpha$ that
makes ${\cal S} $ maximal corresponds to $\alpha = -5.44$, this value
conflicts with  the requirement  $\beta < 4\alpha^2 / 3$ for the
existence of an inflection point. We take instead $\alpha = -3.69$ which
makes ${\cal S} $ the largest possible one and at the same time agrees within
a few percent with approximating $\phi_f\simeq -{\alpha\over 2}$.
Then $\beta \simeq \alpha^2 + y \alpha^6 \simeq 15.92 $. This guarantees
we are effectively exploring the neighbourhood of the curve
$\beta = \alpha^2$ in parameter space.
A plot of the potential for these particular values is given in Fig. 4.
The slow--roll
parameters in this case have values: $\epsilon = 5.94\times 10^{-3}$,
$\eta = 5.84\times 10^{-3}$ and $\xi = 1.32\times 10^{-1}$.
By using Eq.(\ref{12}) we get $n\simeq 0.99$.

Let us consider now the case where $\alpha^2 < \beta < 10\alpha^2 / 9$.
This choice of potential
parameters leads to ${\cal S} < 0$. The non--Gaussian
signal gets maximized for $\alpha = -2.24$ and $\beta = 5.13$.
Fig. 5 shows the potential in this case.
Note the resemblance between this form of the potential and that of the
hybrid \cite{hybrid}
model $V_0 + m^2\phi^2 /2$ for $V_0$ dominating and in the
vicinity of the flat region.
In that case the same sign of the dimensionless skewness \cite{3point}
and blue spectra \cite{blue}, $n>1$, were predicted.
For this value of $\alpha$ we get: $\epsilon = 9.33\times 10^{-4}$,
$\eta = 6.76\times 10^{-3}$ and $\xi = 2.76$.
We get $n\simeq 1.01$.

We can now use Eq.(\ref{quad}) for the quadrupole to find the overall
normalization constant $A$ of the potential.
Bennett et al. \cite{Bennett+Co94} get a best fit
$Q_{{\it rms}-PS} = (17.6\pm 1.5)\mu K$ for $n$ fixed to one, as it is our
present case to a very good approximation.
Thus, for $\alpha = -3.69$ we get $A=2.87\times 10^{-12}$
and ${\cal S}\simeq 1.1\times 10^{-6}$
while for $\alpha = -2.24$ we get $A=5.88\times 10^{-12}$
and ${\cal S}\simeq -1.9\times 10^{-6}$.

The above two values for $n$ show that the departure from scale invariance is
actually very small (in fact, this is because we are
exploring a very narrow range of scales).
Note also the relatively large value of $\xi$ in the
last case compared with $\epsilon$ and $\eta$. This tells us
that the terms proportional to $\xi$ are non negligible in general.
Also the rather small amplitudes for ${\cal S}$ agree with
previous numerical analyses \cite{Hodges+Co90} in which adherence to the
correct level of anisotropies in the CMB radiation under the simplest
assumptions of inflation, like slow--rolling down with potential
(\ref{PolyPot}), practically precludes any observable non--Gaussian
signal.

\section{Conclusions}

In the present paper we have presented explicit calculations
(in the frame of some well motivated inflationary models) of the
dimensionless skewness ${\cal S}$
predicted for the large angular scale temperature
anisotropies in the CMB radiation as well as the evaluation of the
primordial spectral index of the density perturbations originating these
anisotropies.
These computations were performed to full second order in
perturbation theory. In all three models
(even in the case of the polynomial inflaton potential
where more parameters were at our disposal)
very low values for the non--Gaussian signal were obtained.

In fact, the explicit values for ${\cal S}$ were found generically
much smaller than the dimensionless {\em rms} skewness calculated from an
underlying Gaussian density field and are therefore
hidden by this theoretical noise, making experimental detection impossible.

One may try to resort to many--field models in the hope of shifting the
non--zero ${\cal S}$ window to larger values.
In this respect a potential of the form
$V(\phi_i) \sim \exp (\sum_i\lambda_i\phi_i)$
is likely to do well the job \cite{Salopek92}.
In this case the resolution of a set of coupled Langevin--type equations
for the coarse--grained fields (suitably smoothed over a scale larger than
the Hubble radius)  should be faced.
In contrast to previous many--field analyses where one of the fields was
assumed to dominate at a certain stage, in our case (for the aim of
computing the three--point function) we need to make a second--order
perturbative expansion in $\delta\phi_i$ around the classical solutions
$\phi_i^{class}$ but {\rm keeping} $V(\phi_i)$ fully dependent on all the
fields. Non--Gaussian fluctuations can indeed be generated within this
model and thus the prospect of getting a non--negligible value for
the dimensionless skewness should be tested.
This is the subject of our current research.

\vspace{5mm}

\noindent{\bf Acknowledgements:} The author is grateful to
S. Matarrese for valuable discussions. He also thanks A. Conti for his kind
help in the numerical calculations. This work was supported by
the Italian MURST.

\newpage

\section*{Figure Captions}
\begin{description}

\item[Fig. 1 :]  Dimensionless skewness ${\cal S}$
(in units of the quadrupole ${\cal Q}$, see Eq.(\ref{11}))
as a function of $d\equiv \phi^2 - \beta^2 / 16\pi$, the
value of the squared of the field beyond the minimum allowed,
for different choices of $\beta$. Curves from bottom to top
correspond to $\beta$ from 1 to 8 (left panel).

\item[Fig. 2 :]  Spectral index (as calculated from Eq.(\ref{12}))
as a function of $d$. Now curves from top to bottom are those
corresponding to $\beta$ from 1 to 8 (right panel).

\item[Fig. 3 :] Inflaton potential $V(\phi)$ for the parameter choice
$\beta = 1.2$ ($f = 0.77$).
The field is taken in units of $m_P$, while the potential is normalized in
units of $(10^{-1} m_P A^{1/f})^2$.

\item[Fig. 4 :] Polynomial potential $V(\phi)$ as a function of the
scalar field ($\phi$ is taken in Planck mass units). The overall
normalization parameter $A$ is taken to be one;
for $\alpha = -3.69$ and $\beta = 15.92$ (left panel);

\item[Fig. 5 :] Polynomial potential $V(\phi)$ as a function of the
scalar field ($\phi$ is taken in Planck mass units). The overall
normalization parameter $A$ is taken to be one;
for $\alpha = -2.24$ and $\beta =  5.13$ (right panel).

\end{description}


\begin{thebibliography}{99}
\bibitem{f1} In Ref.\cite{3point} an additional contribution to the
	squared brackets in Eq.(\ref{skew}) of the form ${\cal G} = 4 m_P
	\int_{k_*}^{k_{60}} (dq/q) B(\alpha(q)) X(\alpha(q))$
	was found (the scale $k_*$ signals the time when we start to solve
	the Langevin equation, corresponding to a patch of the universe
	homogeneous on a scale slightly above our present horizon).
	In that paper ${\cal G}$ turned out to be always negligible or
	zero for the specific models there considered. As we will see below
	also here
	this term yields a value much smaller than the others explicitly
	written in (\ref{skew}); e.g. within the
	frame of natural inflation ${\cal G}$ can be worked out easily
	yielding  $\vert {\cal G}\vert  \sim 1.1\times 10^{-6} (m_P / f)^4 $,
	much smaller than
	$ X_{60}^2 - 4 m_P X_{60}' \simeq 2 (m_P / f)^2$
	($f\sim m_P$, see below).
\bibitem{f2} 	It is easy to relate these parameters to the steepness
	and its derivatives. In fact
	$\epsilon = X^2 / (16\pi)$,
	$\eta  = \epsilon + m_P X' / (8\pi)$ and
	$\xi  =  \epsilon + 3 m_P X' / (8\pi) + m_P^2 X'' / (4\pi X)$.
\bibitem{f3} 	Recently Kolb and Vadas \cite{KolbVadas94} have given
	second--order accurate expressions for $1-n$, $n_T$ and $R$
	(ratio of the tensor to scalar contributions to density
	perturbations). Applied to our present model this importantly
	changes the relation between $R$ and $1-n$. However the
	second--order correction to $n$
	($\sim 0.2 (m_P^2 / 8\pi f^2)^2\sim 6\times 10^{-3}$ for our
	choice of $f$) is not relevant.
\bibitem{f5} 	Although the slow--rolling assumption is not
	necessary for finding solutions of the  field equations, it is
	instructive to focus also on this limit as it represents the attractor
	solution for large enough time.
\bibitem{f6} 	In this region we can
	effectively neglect the term $\propto\phi^{-2}$
	in the parenthesis of Eq.(\ref{IntPot}) leading to a simplified
	form for $V$.
\end{thebibliography}
\end{document}